\documentstyle[preprint,epsfig]{jpsj}

\title{Spectral Properties of the Chalker-Coddington Network}
\author{Marcus {\sc Metzler}
and Imre {\sc Varga}$^{1,}$\footnote{Permanent address: Department of 
Theoretical Physics, Institut of Physics, Technical University of Budapest,
H-1521 Budapest, Hungary}}
\inst{Department of Physics, Toho University, Miyama 2-2-1, Funabashi, 
Chiba 274\\
$^1$ Institut f\"ur Theoretische Physik, Unversit\"at zu K\"oln, Z\"ulpicher
Str. 77, D-50937 K\"oln, Germany}
\recdate{\today}
\abst{
We numerically investigate the spectral statistics of pseudo-energies 
for the unitary network operator $U$ of the Chalker--Coddington network. 
The shape of the level spacing distribution as well the scaling of its
moments is compared to known results for quantum Hall systems. We also
discuss the influence of multifractality on the tail of the spacing 
distribution.
} 

\kword{
Chalker-Coddington model, level spacing distribution, scaling, 
quantum Hall systems, multifractality}

\begin{document}
\maketitle
\section{Introduction}
\label{sec:intro}
The Chalker-Coddington network model \cite{Chalker_Coddington}, representing
 systems of independent two-dimensional (2D) electrons
in a strong perpendicular magnetic field and smooth disorder potential,
is a convenient tool to investigate the quantum Hall universality class
\cite{Chalker_Coddington,Klesse_Metzler1,Klesse_Metzler2}.
It has been used to determine various quantities characterizing the
localization-delocalization (LD) transition taking place between
the quantized plateaus of the Hall conductance\cite{book}.
The first approach was to use the transfer matrix method to numerically
determine the characteristic exponent $\nu\approx 2.35$ for the 
divergence of the localization length $\xi\propto |E-E_{\rm c}|^{-\nu}$ when
the energy of the electron $E$
approaches the critical energy $E_{\rm c}$ \cite{Chalker_Coddington,LWK}.
At the critical energy itself the network model was used to calculate
critical wave functions. Those were subjected to a multifractal analysis
and the critical exponent $\alpha_0\approx 2.27$ describing the scaling of 
typical value of their squared amplitude with the system size, 
$\exp\langle |\Psi_{\rm c}|^2\rangle\propto L^{-\alpha_0}$,
was determined\cite{Klesse_Metzler1}.
The model can also very conveniently be used to simulate the diffusion of 
wave packets using a unitary time evolution operator $U$. By doing so it
was possible to determine the scaling behavior of the
local density of states \cite{Huckestein_Klesse,Klesse}.

Recently, a new approach in the investigation of network models was
suggested. It was argued that network models can be used to determine the 
statistics of the eigenvalue spectrum of their underlying system
by investigating the spectrum of the unitary network operator
$U$\cite{Klesse_Metzler2}. So far the numerical investigation of the 
pseudo-energy statistics has been concentrating
on the number variance $\Sigma_2(N)=\langle (n-\langle n\rangle)^2\rangle$
of an energy interval containing on average $N=\langle n\rangle$ levels and
the compressibility $\chi=\lim_{N\rightarrow \infty}\lim_{L\rightarrow\infty}
\frac{d\Sigma_2}{dN}$. It was found that the prediction by Chalker et al
\cite{CKL} that $\chi=(d-D_2)/(2d)$, where $D_2$ is the correlation dimension,
can be confirmed. Furthermore, it was seen 
that the level spacing distribution function $P(s)$, denoting the probability 
to find two consecutive levels separated by an energy
$\epsilon=s\Delta$, $\Delta$ being the average level spacing, deviates in 
its tail from the  Wigner surmise \cite{Klesse_Metzler2}. This was expected at
the critical energy \cite{SSSLS,KLAA1,KLAA2} and was also seen in other numerical
investigations\cite{BEO1,BEO2,BEO3}. In this paper we are going to take a closer look
at the level spacing distribution $P(s)$. We will show that by using the 
spectrum of the unitary network operator $U$ all the known results
regarding $P(s)$ can be reproduced. Furthermore, we will introduce an
improved numerical method to determine that spectrum by reducing the
size of the matrix that has to be diagonalized.

%--------------------------------------------------------------------------
\section{The Network Operator and its quasi energy spectrum}
\label{sec:U}
The Chalker-Coddington network is a 2D regular lattice  whose nodes consist
of  elementary scatterers. The scatterers are connected by unidirectional
channels. Each scatterer is a $2\times 2$
matrix $S^k$ mapping 2 incoming channel amplitudes onto 2 outgoing channels.
In order to simulate random channel lengths, a random kinetic phase is
added to each channel amplitude. Vertical
links are always scattered to horizontal links and vice versa. The
scattering probability to the left and the right are determined
by a transmission amplitude $T$ and a reflection amplitude $R=1-T$,
respectively.  
This chiral structure of the network captures the basic features
of the motion of 2D electrons in a strong magnetic field and smooth random
potential, i.e.\ the potential correlation length $d$ is much larger than 
the magnetic length $l$. In this picture the network links represent
the equipotential contours of the random  potential and its nodes are
the saddle points. The scattering amplitude is determined by the
electron energy $E$ and the energy $u_k$ of the respective saddle point, 
$T=(1+\exp((E-u_k)/E_t))^{-1}$, where $E_t$ is the characteristic
tunneling energy of the potential \cite{Fertig,Klesse_Metzler1}.
The saddle point energies can be either fixed ($u_k=0$ for all $k$) or randomly
distributed ($u_i\in [-W,W]$), in any case an LD phase transition occurs
when $E$ approaches the critical energy $E_{\rm c}=0$ \cite{Chalker_Coddington}. 
In the following we assume that all
saddle points have energy $u_k=0$ , i.e.\ $W=0$, corresponding to the
original Chalker-Coddington model. This is done to avoid effects
caused by classical percolation. We will discuss the case $W> 0$ and the
corresponding percolation effects in a later publication.

A state or wave function of the network is defined by a normalized vector 
$\Psi=(\psi_1,\dots,\psi_n)$ whose elements $\psi_i$ denote the complex 
amplitudes on the $n=2L^2$ network channels, where $L$ is the system size.
Its time evolution is determined by the scattering matrices at the saddle
points which can be incorporated into a single unitary network operator $U$:
$$
\psi_i(t+1) = \sum^{2L^2}_j U_{ij}\psi_j(t),
$$
where the $U_{ij}$  are either appropiately arranged matrix elements of the 
scattering matrices $S^k$ or zero. This results in $U$ having only two non-zero
elements per matrix column and being
energy dependent through the energy dependence of the scattering
amplitudes. An energy eigenstate $\Psi$ of the underlying electronic system is
a solution of the stationarity equation:
\begin{equation}
  \label{stationary}
  U(E)\Psi=\Psi,
\end{equation}
which corresponds to an eigenstate of $U(E)$ with eigenvalue 1.
Such an eigenstate will occur only at discrete values of $E=E_n$ which 
form the energy spectrum of the modeled system. In contrast to these
energy eigenvalues, the eigenvalue spectrum of $U(E)$, defined by the 
equation
\begin{equation}
  \label{Ueigen}
  U(E)\Psi_\alpha=e^{i\omega_\alpha(E)}\Psi_\alpha, 
\end{equation}
for a fixed value of $E$ is much easier to determine\cite{Klesse_Metzler1}.
As was pointed out in Ref.~\citen{Klesse_Metzler2}, the eigenphases
$\omega_\alpha(E)$ show the same statistics as eigenenergies $E_n$
close to $E$. They can be considered an excitation spectrum at energy $E$
and we will refer to them as pseudo-energies. 

The fact that we can choose the energy $E$ at which we want to investigate 
the level statistics makes it possible to look directly at the critical
energy $E_{\rm c}$ without having to worry about leaving the critical region.
This means that all the critical pseudo-energies, i.e. the eigenvalues of 
$U(E_{\rm c})$, can be used for our statistics, which is an enormous advantage over
other methods that can only use part of the eigenspectrum of their 
Hamiltonians. In order to further improve the efficiency of the numerics,
we also use the fact that for any eigenphase $\omega_\alpha$
the phase $\omega_\alpha+\pi$ is also an eigenphase. 
This pseudo-degeneracy is caused by the chiral structure of the 
Chalker--Coddington model and the resulting fact that each time step
maps horizontal to vertical links only and vice versa. If you look at $U^2$
this property results in the formation of two orthogonal $U^2$-invariant 
subspaces that have the same eigenvalues. This $U^2$ degeneracy is reflected
in the pseudo-degeneracy in $U$, which
forced us to disregard half the eigenvalues in the past. Now we will
use it to reduce the size of our matrix. When  diagonalizing the  
matrix $U^2$, instead of $U$, and calculating the degenerate eigenphases 
$2\omega_\alpha$, we can take out all matrix elements corresponding
to vertical (or horizontal) links 
and thereby reduce the matrix to a quarter of its size.
Diagonalizing this stripped matrix and dividing all its eigenphases
by 2 gives us the upper semi-circle of all eigenvalues of $U$, which
is the maximum yield possible, within a much shorter time.
Now let us take a look at the results.

%----------------------------------------------------------------------
\section{The level spacing distribution $P(s)$ and scaling}
\label{sec:Pofs}
The level spacing distribution in the localized regime, i.e.\ at energies
$|E|\gg E_{\rm c}=0$ is Poissonian: $P_P(s)=\exp(-s)$. The closer we get
to the critical energy the more it will change and become closer
to the GUE distribution of the metallic phase known from
random matrix theory\cite{rmt}, 
$P_{GUE}(s)=\frac{32}{\pi^2}s^2\exp\left(-\frac{4}{\pi}s^2\right)$
(see Fig.~\ref{fig2}).
Since we do not have a metallic phase in the quantum Hall regime, the
GUE distribution will not be reached. At the critical energy the distribution
is neither GUE nor Poisson, because the critical wave functions are neither
smeared out homogeneously nor are they localized on a small area. Instead, 
they form a self similar measure fluctuating  strongly on all length scales, 
which is best described in terms of multifractality\cite{Janssen}.
It has been found that at criticality the $P(s)$ distribution is best 
overall fitted by the following formula\cite{BEO3}:
\begin{equation}
  \label{Posdist}
  P(s)=As^\beta\exp\left(-Bs^\alpha\right),
\end{equation}
where $A$ and $B$ are determined by the
normalization conditions $\langle 1\rangle=1$ and $\langle s\rangle=1$.
This shape was proposed by analytical theories that predicted that
$\alpha=1+1/(d\nu)\approx 1.21$\cite{KLAA1,KLAA2}. We will see that this prediction
is not supported by our data, instead an earlier prediction \cite{AZKS} for 
the tail region of $P(s)$ that makes a connection to the compressibility
$\chi$ can be verified, i.e. $P(s)\propto \exp(-s/(2\chi))$.
 
In our investigation we determined pseudo-energy spectra for system
sizes $L$ ranging from 16 to 50, i.e.\ $L^2$ saddle points and
eigenvalues per system. The number of realizations per system size
was set to get from $4\times 10^4$ up to $5\times 10^6$ eigenvalues. 
This resulted in enough statistics to get good results for $P(s)$ 
for values of $s$ up to about $s=4.0$. In Fig.~\ref{fig2} $P(s)$ is 
shown at different energies ranging from $E=E_{\rm c}=0$ to $E=3$. The change 
from the critical distribution at $E=0$, which does not fit the GUE 
behavior, to the Poissonian type, which is almost reached at $E=3$, 
can be seen very clearly. 
This change is expected to
show one-parameter scaling for any function $Z(E,L)$ that describes the
shape of $P(s)$:
\begin{equation}
  \label{scaling}
  Z(E,L)=f(L/\xi),\quad \xi=\xi_0 |E-E_{\rm c}|^{-\nu}.
\end{equation}
As a scaling variable we consider several quantities, first \cite{ZhKr} 
\begin{equation}
  \label{j0}
  J_0(L,E)=\int_0^\infty Q_0(s) ds=\frac{1}{2}\int_0^\infty s^2 P(s)ds,  
\end{equation}
where $Q_0(s)$ is the probability that an energy interval of width $s$ 
contains no energy eigenvalue. Note that $J_0$ is half the lowest order 
momentum of $P(s)$ that is nontrivial and is also connected to 
$q=\langle s^2\rangle ^{-1}$ that describes the
`peakedness' of a distribution\cite{VHSP,VOOP}. Furthermore, 
we invoke the entropic 
moment of $P(s)$, $-\langle s\ln(s)\rangle $ and calculate, apart 
from $-\ln(q)$, another generalized R\'enyi--entropy 
$S_{\rm str}=-\langle s\ln(s)\rangle +\ln\langle s^2\rangle$ 
\cite{VHSP,VOOP,VPJP}.

All of these quantities are numerically easily obtainable and they
take all calculated neighboring level spacings into account. Since all of
them show nice scaling behavior we present only that of $J_0$.
The inset in Fig.~\ref{fig3} shows $J_0$ as a function of $E$ for different
system sizes. As expected, all the curves meet in a single size independent
point $J_0^c=0.61\pm 0.01$ at the critical energy $E=E_{\rm c}=0$.
If we assume the validity of Eq.~(\ref{scaling}) we can rescale $L$ so that
the data points of $J_0$ as a function of $L/\xi$ for consecutive energies $E$
overlap with each other. The result of such a rescaling is shown in 
Fig.~\ref{fig3}. In this case we fit a fourth degree polynomial in 
$\ln (L/\xi_0)+\nu\ln E$ to a log-log plot of our data and thereby determined
$\nu=2.1\pm 0.3$. This is consistent with other results for $\nu$
obtained by various previous studies\cite{nu1,nu2,nu3,nu4,nu5}.

Now, let us proceed with the shape analysis of the $P(s)$ function using the 
other two quantities $q$ and $S_{\rm str}$. As it has been shown in a number of 
cases \cite{VHSP,VPJP} these parameters apart from showing scaling and fixed 
point in the critical regime, as a bonus enable us to determine the 
one--parameter family of $P(s)$ functions describing the transition from e.g.
GUE-like to Poissonian behavior. To this end we apply the same linear 
rescaling as was done in Ref~\citen{VOOP} and introduce $\tilde{Q}$ and 
$\tilde{S}$ that span the whole $[0,1]$ interval with $0$ $(1)$ corresponding 
to the Poissonian (GUE-like) limits, respectively. In Fig.~\ref{fig4} the 
parametric dependence of $\tilde{S}$ vs $\tilde{Q}$ is shown. We clearly see, 
that all the data fall on a common curve as was already seen in numerical 
simulation of the random--Landau matrix model \cite{VOOP}. Furthermore, as a
comparison, the behavior of the $\tilde{S}(\tilde{Q})$ dependence is also 
given for three families of $P(s)$ functions based on 
Eq.~(\ref{Posdist}): $a$) fixed $\beta=2$ and variable $\alpha$ in the range
$[1,2]$; $b$) fixed $\alpha=1$ and variable $\beta$ in the range $[0,2]$; and
$c$) using $\beta=2\gamma$ and $\alpha=\gamma +1$ with variable $\gamma$ in the
range $[0,1]$. The first and the second families meat each other at the
``semi--Poissonian'' of the form $P(s)=\frac{27}{2}s^2\exp(-3s)$ similar to
that proposed in Ref~\citen{SSSLS}. The family $c$ is a Brody--like form that has 
already been seen to fit the evolution of the system in the 
$\tilde{S}$ vs $\tilde{Q}$ parameter space in Ref~\citen{VOOP}. As we can see our
data indicate a similar behavior as the random--Landau matrix model. This is
another nice example of the presence of one--parameter scaling since all data
can be described by a one--parameter family of $P(s)$ functions. This analysis
suggests $\gamma\approx 0.54$ around the critical point ($E=0$).

Right at the critical energy, however, the two parameter fit of $P(s)$ to 
Eq.~(\ref{Posdist}) is shown in Fig.~\ref{fig5}. The fit was done for all 
available system sizes and resulted in the averaged values 
$\alpha=1.56\pm 0.04$ and $\beta=1.99\pm 0.03$. The errors were determined 
by the fitting and averaging procedures. Within those errors the result is 
almost the same as that obtained by Ono {\it et al}. for the Landau model in 
the lowest Landau band \cite{BEO3} ($\alpha=1.53$, $\beta=2.19$). 
This contradicts the prediction that $\alpha=1.21$ \cite{KLAA1,KLAA2}.

In Ref~\citen{AZKS} it was argued that the tail should show an exponetial decay.
The results of a fit to a function $\exp(-\kappa s)$ are shown in the 
inset of Fig.~\ref{fig5}. We get $\kappa=3.95\pm 0.6$, which is comparable
to the result by Batsch and Schweitzer\cite{BS} ($\kappa\approx 4.1$) and
fits the prediction $\kappa=1/(2\chi)$, with 
$\chi = 0.124\pm 0.06$ \cite{Klesse_Metzler2}.

%---------------------------------------------------------------------
\section{Conclusion}
In conclusion we have presented the statistical analysis of spectra
of pseudo-energies taken at different electron energies obtained in 
a network--model of the quantum--Hall effect. We could confirm
that those pseudo-energy spectra show the same statistics as those of the
energy eigenvalues of other models of the quantum Hall systems. This was
done by investigating the shape of the level spacing distribution function
$P(s)$ and the scaling of its moments $J_0=\langle s^2\rangle /2$, 
$q=\langle s^2\rangle ^{-1}$ and $-\langle s\ln s\rangle$. Furthermore, 
although the overall shape of $P(s)$ at the critical point has yet to be
explained, we observe that level repulsion leads to the same 
quadratic behavior for small $s$ as in the metallic regime, whereas the 
tail of the $P(s)$ distribution falls off more slowly than in the GUE case. 
This is due to the multifractality of the critical wave functions and is 
also reflected in the finite compressibility ($0<\chi<1$). 

%----------------------------------------------------------------------
\section*{Acknowledgements}
We would like to thank M. Janssen and Y. Ono for valuable discussions.
Partial support from DFG and SFB-341 as well as OTKA Nos. T021228, 
T024136, F024135 are gratefully acknowledged.

\newpage

\begin{figure}
  \begin{center}
  \end{center}
  \caption{The $P(s)$ spectrum for energies ranging from $E=0$ to
    $E=3$. The smooth curves show the behavior for Poisson and GUE.
     The inset shows the same in double logarithmic plot.}
    \label{fig2}
\end{figure}

\begin{figure}
  \begin{center}
  \end{center}
  \caption{The one-parameter dependence of the scaling variable $J_0$ on 
    $L/\xi(E)$. The inset shows $J_0$ as a function of energy $E$ for
    different system sizes. At the critical point $E_{\rm c}=0$ all the curves
    meet at the same point.}
    \label{fig3}
\end{figure}

\begin{figure}
  \begin{center}
  \end{center}
  \caption{The parametric dependence of $\tilde{S}$ vs $\tilde{Q}$ for
    our numerical results compared to three families of $P(s)$ functions
    as described in the text.}
    \label{fig4}
\end{figure}

\begin{figure}
  \begin{center}
  \end{center}
  \caption{The $P(s)$ distribution for $E=E_{\rm c}=0$. We can see the data
    for $L=20, 24, 30, 40, 50$. All data points are positioned on the 
    smooth curve that is a fit to Eq.~(\ref{Posdist}), where
    $\alpha=1.56\pm 0.04$ and $\beta=1.99\pm 0.03$. The inset shows a fit of
     $\exp(-\kappa s)$ to the data for $L=40$, $\kappa=3.95\pm 0.06$. }
    \label{fig5}
\end{figure}

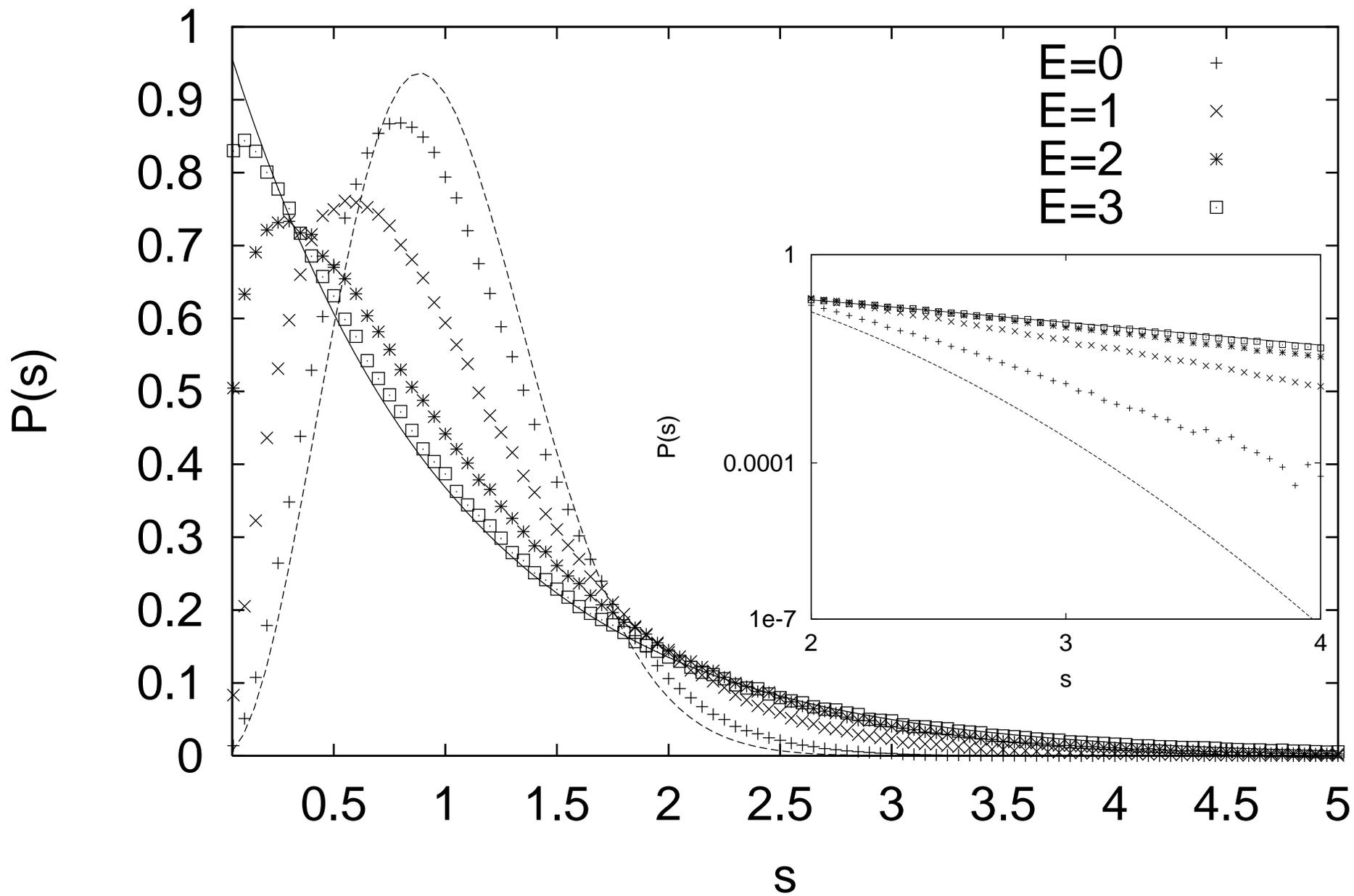
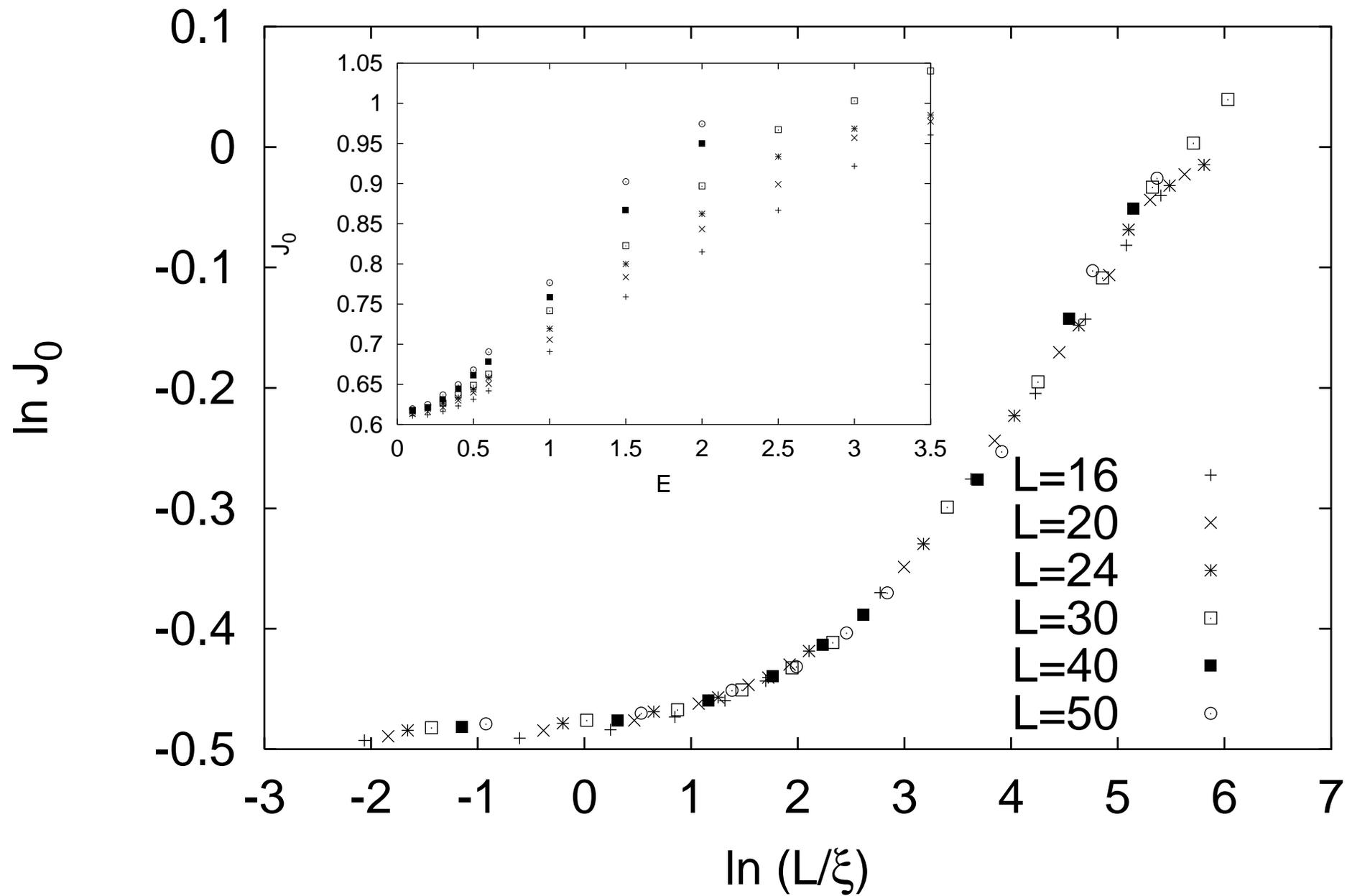
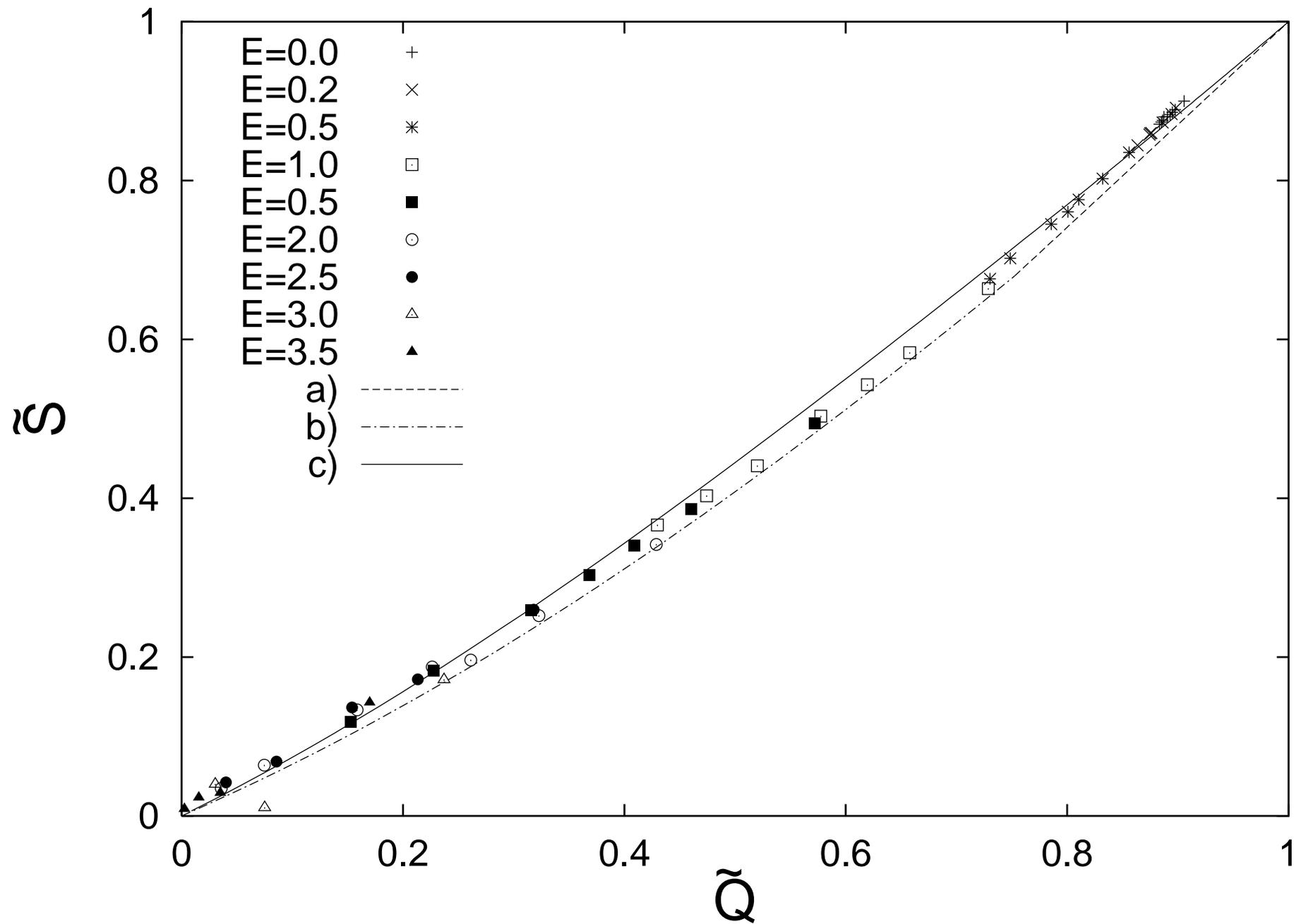
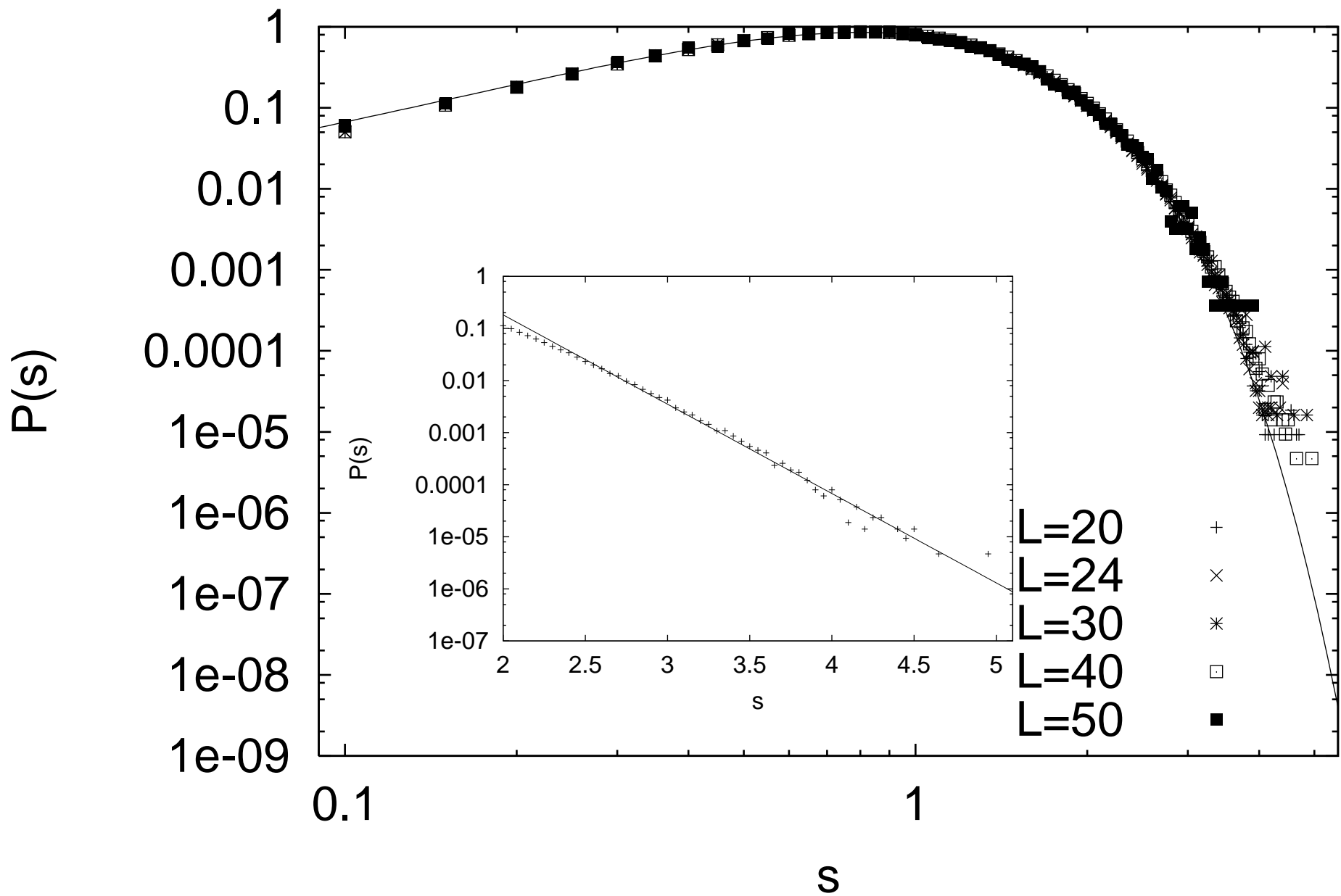
\end{document}